\documentclass[useAMS]{mn2e_modmargin}

\usepackage{amsmath}
\usepackage{url}
\usepackage{amsfonts}
\usepackage{amsbsy}
\usepackage{graphicx}
\usepackage{subfigure}
\usepackage{verbatim}
\usepackage{amssymb}

\newcommand{\ex}{\ensuremath{\mathbf{e}_{x}}}
\newcommand{\ey}{\ensuremath{\mathbf{e}_{y}}}
\newcommand{\ez}{\ensuremath{\mathbf{e}_{z}}}

\title[The dynamics of a stream of co-orbital particles]
{The gravitational instability of a stream of co-orbital particles}

\author[Henrik N. Latter]{Henrik N. Latter$^{1}$\thanks{Email:
hl278@cam.ac.uk}, Hanno Rein$^{2}$,
 Gordon I. Ogilvie$^{1}$
\\ \\
$^{1}$ DAMTP, University of Cambridge, CMS, Wilberforce Rd, Cambridge
CB3 0WA, UK \\
$^{2}$ Institute for Advanced Study, 1 Einstein Dr, Princeton, New
Jersey 08540, USA}

\begin{document}
 
\maketitle

\label{firstpage}

\begin{abstract}
We describe the dynamics of a stream of equally spaced macroscopic 
particles
in orbit around a central body (e.g.\ a planet or star). A co-orbital
configuration of small bodies may be subject to
gravitational instability, which takes the system to
 a spreading disordered and collisional state. We detail the linear instability's
mathematical and physical features using the shearing sheet model 
and subsequently track its nonlinear evolution with
local N-body simulations. This model provides a convenient tool with
which to understand the
gravitational and collisional dynamics of narrow belts, such as Saturn's
F-ring and the streams of material wrenched from tidally
disrupted bodies. In particular,
we study the tendency of these systems to form long-lived particle
aggregates. Finally, we uncover an unexpected connection between the linear
dynamics of the gravitational instability and the magnetorotational instability.
\end{abstract}

\begin{keywords}
   instabilities --- planets and satellites: dynamical evolution and
   stability, rings 
\end{keywords}

\section{Introduction}

This paper is concerned with the gravitational and collisional
 dynamics of belts of material orbiting planets or stars.
 Of particular interest are Saturn's F-ring
and the tidal streams torn from disrupted satellites. 
The F-ring is thought to comprise a population of large
objects (of some 10 km) swathed in dust (Showalter 2004, Esposito et
al.~2008, Murray et al.~2008). Being located so near the Roche limit, the
size distribution of these larger bodies evolves
 according to gravitational
aggregation, and tidal and collisional disruption (Barbara and
Esposito 2002, Esposito et al.~2012). The tidal environment
is very different for those dense narrow rings located interior the
Roche limit, such as the $\epsilon$-ring of Uranus or the dense
ringlets ensconced in Saturn's C-ring (Colwell et al.~2009). It is possible that these dense
ringlets originated from the tidal disruption of 
 small satellites (or the mantles thereof), with their early
evolution controlled by gravitational instabilities
(Leinhardt et al.~2012). 

We would like to understand these
several processes --- tidal disruption, particle clumping,
gravitational instability, collisions --- in a single theoretical
 framework using a simple but illuminating analytical model and
 N-body simulations. The main configuration that we consider is
 the gravitational stability of a stream, or line, of many
 equally spaced particles. To facilitate its study, we employ a
 local model: the shearing sheet. In doing so we can more 
easily reproduce the famous result of Maxwell's Adams Prize essay
(Maxwell 1859, Cook and Franklin 1964) and form a more intuitive
understanding of the instability's physics. We find a
stream of particles is susceptible to two kinds of disruption: a
familiar gravitational clumping instability, and a growing epicyclic
mode. Which instability the system selects depends on
 a single dimensionless parameter analogous to the
Roche parameter that controls the tidal disruption of a satellite. 

The
nonlinear outcome of these instabilities is studied with N-body
simulations using the code REBOUND (Rein and Liu 2012).
We find that the precise form of the
initial instability is unimportant; instead, 
 the nonlinear evolution is entirely controlled by the tendency
of the system to form particle aggregates.
 Existing clustering
criteria are discussed (Weidenschilling et al.~1984, Ohtsuki
1993, Canup and Esposito 1995), and we 
simulate an interesting intermediate regime relevant to
Saturn's outer rings, whereby
collisional aggregation is rare, but not impossible, and instead
particles form temporary short-lived clusters 
(cf.\ `dynamical ephemeral bodies', Weidenschilling et al.~1984).

These gravitational instabilities may illuminate other astrophysical contexts. For
instance, the mechanism of instability we outline here could also be at
work in the non-axisymmetric instabilities of narrow \emph{fluid} rings and
slender tori (Maxwell 1859, Goodman and Narayan 1988, Papaloizou and
Lin 1989). 
It also manifests in the
`propeller and frog' resonance
 of Pan and Chiang (2010, 2012) which aims to describe the
observed migration of large embedded objects (propellers)
 in Saturn's A-ring. If the
propeller is permitted to back-react on the rest of the ring, then an
instability will arise, analogous to the one we study here. 
Lastly we show how the mechanism of instability in the growing
epicyclic mode, relying on significant angular
momentum exchange, shares some of the mathematical and physical
characteristics of the magnetorotational
instability (MRI) (Balbus and Hawley 1991). In fact, 
the two instabilities are nearly identical for a certain
(somewhat artificial) equilibrium set-up.

The following section outlines the linear theory of the gravitational
instability, connecting it to previous work by Maxwell (1859) and Fermi and
Chandrasekhar (1953), while showing its similarities to the MRI\footnote{At the conclusion of this work we discovered that some of the
linear theory we present in Section 2 was treated in a similar way by Willerding (1986),
though our emphases and interpretations differ.}. Section
3 presents N-body simulations of the instability's nonlinear and 
collisional evolution. In Section 4 we summarise our results.

\section{Linear Theory}
\subsection{Governing equations and setup}

Consider a train of spherical particles in a circular orbit around a
point mass or around a central mass with a point-mass potential. 
The particles initially inhabit the same orbital radius but are equally
spaced in azimuth by a distance $h$. 
Each particle has the same
mass $m$ which is much smaller than the central mass. 
If the spacing $h$ is sufficiently small and the ensuing
dynamics remain confined to lengthscales much less than the orbital
radius,
 we may adopt the
shearing sheet model or local approximation (Goldreich and Lynden-Bell 1965),
 in which case we can write down the equations of
motion for each particle in a convenient way. The shearing sheet is
anchored at a radius $R_0$ from the central object around which it
 orbits with frequency $\Omega$. In this model, the motion of 
particle $n$ is governed
by
\begin{align} \label{nleq1}
\ddot x_n - 2\Omega \dot y_n &= 3\Omega^2\,x_n + f_x^{n}, \\
\ddot y_n + 2\Omega \dot x_n &= f_y^{n}, \label{nleq2} \\
\ddot z_n &= -\Omega^2\,z_n + f_z^{n}, \label{nleq3}
\end{align}
where $(x_n,\,y_n,\,z_n)$ denotes the coordinates of particle $n$ in the
shearing box relative to the reference position $R_0$, 
an overdot indicates a time derivative, and $\mathbf{f}^n$ is the
specific collective gravitational force of all the other particles on particle
$n$. 
The
gravitational force may be written as
\begin{equation}
\mathbf{f}^n = G\,m\sum_{j\neq n}
\frac{\Delta x_{jn}\ex +
  \Delta y_{jn}\ey +\Delta z_{jn} \ez}
{[\Delta x_{jn}^2 + \Delta y_{jn}^2 + \Delta z_{jn}^2]^{3/2}},
\end{equation}
where the sum is over all other particles,
 with $G$ as the gravitational constant, and the distances 
$$ \Delta x_{jn} = x_j - x_n, \quad \Delta y_{jn} =y_j-y_n, \quad
\Delta z_{jn} = z_j - z_n. $$
In reality there are a large but finite number of particles in the
ring, but in practice we take the above sum to be infinite and so $j$
runs between $-\infty$ and $\infty$.
 Even if the force is approximated as an infinite sum, it
remains convergent because
only nearby particles significantly influence any given particle's dynamics.
 See Salo and Yoder (1988) or Vanderbei
and Koleman (2007) for a
discussion of situations in which the rings contain
only a few massive particles.

\subsection{Equilibrium}

This set of equations admits as an equilibrium state a procession of
equally spaced particles located in the central axis of the sheet, i.e. for
\begin{align} \label{Eqm}
x_n =0, \quad y_n = hn, \quad z_n=0, \qquad \forall\,n \in \mathbb{Z},
\end{align}
where $h$ is the azimuthal spacing. The force $\mathbf{f}^n$ is zero
for each $n$: the gravitational influences of the
particles preceding a given particle $n$ are cancelled exactly by all the
particles trailing particle $n$.

\subsection{Linearised equations}

We next perturb this stream of particles by a set of small
displacements $(x_n',\,y_n')$ that lie only in the orbital plane. 
Vertical displacements are neglected as they decouple
in the linear regime and do not lead to instability.
The linearised equations for these time-dependent
displacements are:
\begin{align}
\ddot x_n' - 2\Omega \dot y_n' - 3\Omega^2 x_n' =
\frac{Gm}{h^3}\sum_{j=-\infty}^\infty \frac{x_j'-x_n'}{|j-n|^3}, \\
\ddot y_n' + 2 \Omega \dot x_n' =
-2\frac{Gm}{h^3}\sum_{j=-\infty}^\infty \frac{y'_j-y'_n}{|j-n|^3}\,,
\end{align}
where both sums omit the $j=n$ term. 

We now suppose the particle motions 
assume a collective Fourier mode of the
type
$$ x_n' = X\,\text{e}^{st+nh k \text{i}}, \qquad y_n' = Y\,\text{e}^{st+n h k \text{i}},
$$
with $X$ and $Y$ complex amplitudes, $s$ a growth rate (complex in general), and $k$
 a (real) wavenumber. In order to avoid aliasing, the magnitude of $k$ is restricted to be
equal or less than $\pi/h$.
 The equations reduce to two,
\begin{align} \label{lin1}
&s^2 X - 2\Omega s Y - 3 \Omega^2 X = -\frac{Gm}{h^3}F\,X, \\
& s^2 Y + 2\Omega s X = 2\frac{Gm}{h^3} F\,Y, \label{lin2}
\end{align}
in which we have introduced the dimensionless 
force function $F$. It can be
manipulated into
\begin{equation} \label{Force}
F(K) = 2\sum_{j=1}^\infty \frac{1-\cos(j\,K)}{j^3},
\end{equation}
where we have introduced a dimensionless wavenumber $K=hk$.
The force function depends only on $K$, and can be re-expressed in
terms of the zeta and Clausen functions (Lewin 1981).
 We find that $F>0$ and is an increasing function of $K$, with $F\to 0 $ as
 $K\to 0$. The following asymptotic
 expansion, for small $K$, offers a reasonable
 approximation for all physical wavenumbers:
$$ F = \left(\frac{3}{2} - \ln K \right)\,K^2 + \frac{1}{96}\,K^4+ \mathcal{O}(K^6),$$
(for a derivation see, for example, Nicorovici et al. 1994). 

\subsection{Springs, clumping, and angular momentum exchange}

Before moving on to the dispersion relation itself, we make some
preliminary remarks.
First, if the central object and the orbital motion are neglected, the
terms involving $\Omega$
in Eqs \eqref{lin1} and \eqref{lin2} disappear. Consequently, the azimuthal and radial
motions decouple and the
azimuthal equation gives rise to an unstable longitudinal mode which
tends to clump particles together. Its growth rate is clear from 
inspection and is equal to
$\sqrt{2GmF/h^3}$. This unstable mode, which grows as the result of a
release of gravitational potential energy, is the discrete analogue of the
instability of an infinite self-gravitating cylinder described by
Chandraskehar and Fermi (1953) (see also Chandrasekhar 1981).

Second, \eqref{lin1} and \eqref{lin2} are
 reminiscent of the equations governing two
orbiting bodies attached by a spring, as popularised by
Balbus in his interpretation of the magnetorotational instability
(Balbus and Hawley 1991). There is, however, a crucial
 difference in that our
`gravitational spring constant' 
 differs in both size \emph{and sign} depending on the direction of the spring
 force.
 While 
 the gravitational 
`spring' seeks to resist extension in the radial direction (as in the MRI),
it actively encourages it in the longitudinal direction. In some sense, the
spring wants to become `clumpy'. At the same time
 the spring forces also facilitate angular momentum exchange,
 which introduces a separate route to instability, employed by the MRI and the
Papaloizou-Pringle instability (Papaloizou and Pringle 1984). Outward
angular momentum transfer liberates orbital energy which is channeled into
exponential growth.
The
gravitational instabilities we study in this paper are the outcome of
the interplay, and sometimes competition, between these two
instability mechanisms.

\subsection{Dispersion relation}

We rescale time so that $\Omega=1$. 
Eliminating $X$ and $Y$ from Eqs \eqref{lin1} and \eqref{lin2}
 yields the following
dispersion relation for $s$:
\begin{equation}\label{disp}
s^4 + [1 - g F(K)]s^2 + 2 g F(K)[3-g F(K)] =0,
\end{equation}
where we have introduced the important dimensionless quantity
\begin{equation}
g = \frac{(m/h^3)\,G}{\Omega^2}.
\end{equation}
It measures the influence of the stream's self-gravity relative to
its inertial forces.

The $g$ parameter bears a superficial resemblance to
 the `Roche parameter' of an orbiting
fluid satellite vulnerable to tidal disruption:
\begin{equation}\label{Roche}
 \mathcal{R}= \pi \frac{\rho\, G}{\Omega^2},
\end{equation}
where $\rho$ is the mass density of the
satellite. 
We find gravitational instability favours
larger $g$ --- hence greater mass densities and greater radii, $R_0$.
In contrast, a larger $\mathcal{R}$ in the Roche problem corresponds
to a
greater resistance to disruption (Chandrasekhar 1987). 
This reflects the fact that
the inertial forces are stabilising in the ring context and disruptive
in the satellite context. As we show later, however, the parameter $g$
does not control the nonlinear saturation of gravitational instability.

Finally, note that by making the identification $g F(K) = v_A^2
k^2/\Omega^2$, where $v_A$  denotes Alfv\'en speed,
Eq.~\eqref{disp} becomes strikingly similar to the MRI dispersion
relation (e.g. Eq.~(79) in Balbus 2003). But differences in sign occur at key places, which we
associate with the tendency to azimuthally clump. 

\subsection{Fastest growing modes and stability criterion}

In this problem the shortest modes are the most unstable: they always
grow the fastest and they are always the first to be
destabilised. Therefore, our stability considerations need only
make reference to these short scales.
The
shortest modes possess $K=\pi$ and the force function is, consequently,
$F(\pi)= (7/2)\zeta(3)\approx 4.2$, where $\zeta(3)$ is Ap\'ery's
constant. From the explicit solution to the (bi-) quadratic \eqref{disp}, we find three
bifurcations as $g$ varies. In order of increasing $g$, these occur at
\begin{equation}
g = \frac{26 \pm 8\sqrt{10}}{63\,\zeta(3)}\approx 0.00927,\,0.677,
\end{equation}
and
\begin{equation}
g= \frac{6}{7\zeta(3)} \approx 0.714.
\end{equation}

The first bifurcation corresponds to the onset of instability.
When
 $g<0.00927$ there exist only neutral modes, and the system is stable.
 Two of the modes in this regime consist of
near-epicyclic motion with frequency $|s| \lesssim \Omega$.
 As $g$ is increased through $0.00927$, there is a Hopf
 bifurcation and
both these
oscillations pick up equal and positive growth rates 
(in addition, there are
decaying modes as the system is Hamiltonian). The gravitational instability
 takes the rather novel form of a pair of growing epicycles (as in the
 viscous overstability --- Borderies et al.~1985, Schmit and
 Tscharnuter 1995, Latter and Ogilvie 2009). 
Note that the stability criterion, $g<0.00927$,
can be reformulated into the
following:
\begin{equation} \label{Maxwell}
\frac{m}{M} < 2.298657/N^3,
\end{equation} 
where $N$ is the total number of particles and $M$ is the mass of the
central object. Here we have defined the number of particles through 
$Nh=2\pi R_0$. Equation \eqref{Maxwell} is
the famous stability criterion that Maxwell derived
 in his Adams Prize essay for
the stability of $N$ co-orbital particles for large $N$ (Maxwell 1859, Cook
and Franklin 1964).  

Now, when $g$ is increased towards $0.677$, the next critical value, the
oscillation frequencies of the two unstable modes are gradually 
suppressed by the self-gravity. At $g=0.677$ these oscillation
frequencies are precisely zero, and there exists a double root to
\eqref{disp}. When $g>0.677$ both unstable modes are monotonically
growing, and cease to oscillate. As $g$ is increased further, these two modes
decouple: one growing more quickly than the other. Eventually, when
$g>0.714$ (the third bifurcation), the slower of the two becomes
neutral, leaving only one growing mode.
The instability now takes the more familiar form of
 longitudinal clumping. This is especially clear
in the limit of large $g$, where the unstable mode's growth rate
approaches $\sqrt{2g\,F(\pi)}$, which is the value for a non-orbiting
line of self-gravitating particles (cf.\ Section 2.4) .

\subsection{Modes of general wavelength}

Modes on longer scales (smaller $K$) follow the same pattern, though the critical
values of $g$ are larger, and are different for each $K$. Note that
sufficiently long modes are \emph{always stable} for any given $g$. 
Once we stipulate $g$ --- no matter how large --- formally we can always find a
sufficiently small
$K$ so that $gF(K)\ll 1$. (In reality, of course, $K$ is limited by
the circumference of the ring.) 
Expanding \eqref{disp} in this small
combination gives the growth rates:
\begin{equation}
 s = \pm i \left[ \sqrt{6g F} +\mathcal{O}(g F)\right], 
\quad s=\pm i\left[1+\mathcal{O}(g F)\right].
\end{equation}
The instability scale is tied to the strength of 
self-gravity. Conversely, in the MRI context, once the strength of the imposed
magnetic field is specified (via the Alfv\'en speed),
 we may always find a wavelength
above which the MRI operates. In the gravitational instability,
 once the strength of the particles'
self-gravity is specified, we may always find a wavelength \emph{below} which the
GI operates (but only when $g>0.00927$).

\subsection{Growth rates}

\begin{figure}
\scalebox{0.5}{\includegraphics{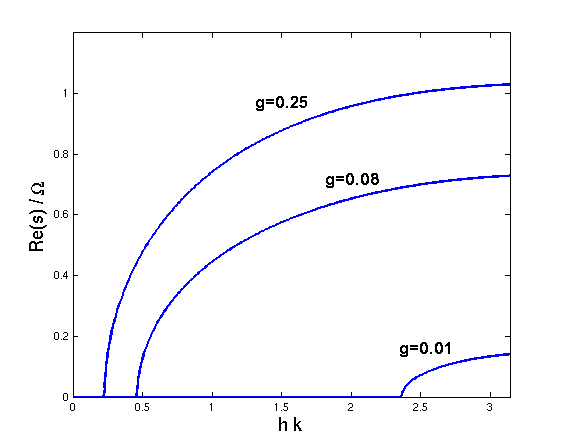}}
 \caption{Growth rates of the gravitational instability as a function
   of dimensionless wavenumber $K$ for three different
   $g=$ 0.01, 0.08, and 0.25.
 Each curve is labelled with its corresponding value of $g$.
  There are two unstable modes with the same
 growth rate and both oscillating with a frequency $\approx \pm
 \Omega$.}
\end{figure}

In Fig.~1, we plot the real part of the growth rate $s$ of the
unstable modes as a function
of $K$, for various $g$. In this figure we concentrate on the
growing epicycles; thus
we limit $g$ to be less than
$g< 0.677$. Three values of $g$ are chosen: 0.01, 0.08,
and 0.25. As explained earlier, the shortest modes (largest $K$) modes
are the most unstable, but $K \leq \pi$. Note
 finally that for sufficiently long modes instability is quenched in
 all three cases.

\begin{figure}
\scalebox{0.5}{\includegraphics{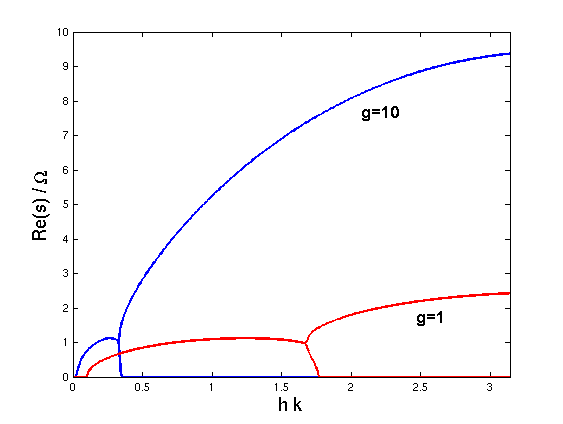}}
 \caption{Growth rates as functions of $K$ for larger values of
   $g$. Two values are chosen: $g=1,\,10$. Above a critical $K$, in
   both cases, there exists only one monotonically growing
   mode. For $K$ below a second critical value, there are two unstable epicyclic
   modes growing at the same rate. This value occurs at the `branches' in the dispersion
   relation. For $K$ less than a third critical value the modes are stable.}
\end{figure}

In Fig.~2, we extend the previous plot to larger $g$, in which we find
the instability taking the form of gravitational clumping on
shorter scales. Two values of $g$ are chosen above the value
$0.677$. In both cases instability is quenched for small $K$.
 One may also 
observe the bifurcation at a second critical $K$ at which point the
instability changes from being two growing epicycles, to two monotonic
clumping modes. For $K$ greater than a slightly larger value, one of
these modes stabilises, and only one growing mode remains.

\subsection{Physical interpretation}

\begin{figure}
\scalebox{0.65}{\includegraphics[angle=270]{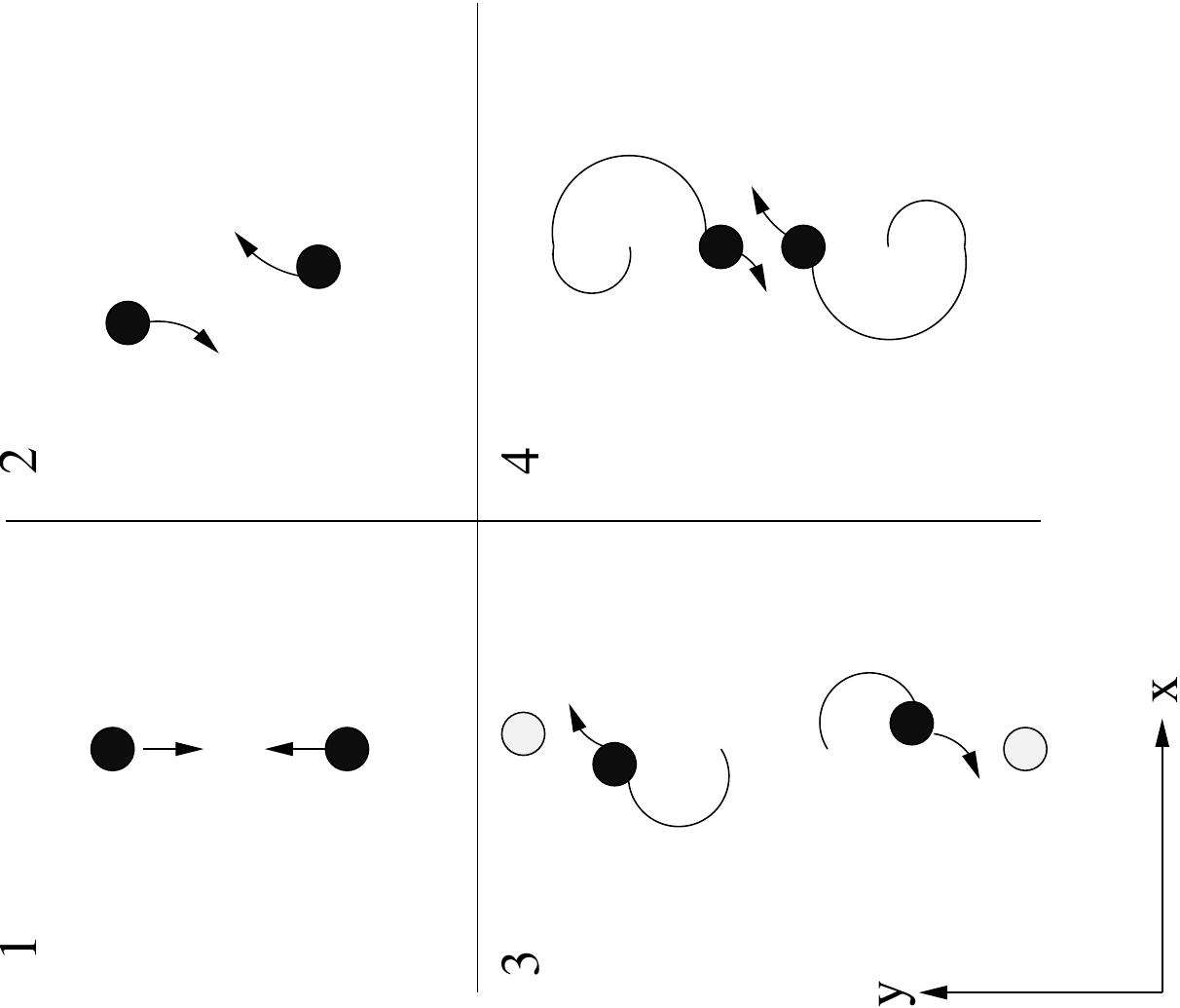}}
 \caption{A cartoon of the onset of oscillatory instability. The first
 panel shows two adjacent particles given a small displacement toward
 one another. In the second, their mutual
 attraction gives rise to an epicyclic motion (akin to a `horseshoe'
 turn), 
the amplitude of which
 is increased by a gravitational encounter with the particles on
 either side of the pair, shown in the third panel. In the last panel,
the two particles kick each other once again giving rise to an even
large (modified) epicyclic motion.}
\end{figure}

The instability associated with longitudinal clumping, at larger $g$,
is relatively straightforward to understand. A pattern of compression
and rarefaction is self-amplifying because gravitational
attraction increases when neighbouring particles are
 displaced closer to each other. In contrast,
 the `epicyclic' instability, which occurs
 for smaller values of $g$, relies on a mechanism that is a little
 more subtle and which we now describe using a simple cartoon.

Consider the four `snapshots' of Fig.~3. In the first panel, we
have drawn just two particles within the initial equilibrium
state. These two particles have been displaced slightly towards each
other (and away from their other immediate neighbours).
As a result, there is a net gravitational attraction between them
(the black arrows). This force `slows down' the leading
particle and `speeds up' the trailing particle; thus an angular momentum
exchange takes place. Consequently, both particles begin an
epicycle  around the two new radii that are
 associated with their new angular momenta: the leading particle
 moves to a smaller radius, while the trailing moves to a larger radius (Panel
 2). Note that if the gravitational attraction is sufficiently strong
 (large $g$) then the particles would clump or collide before the
 epicyclic motion gets properly underway. Panel 3 describes what
 happens when the two particles complete half of their epicycle. At
 this stage they are closer to their other near neighbours (in white) 
than to each other. Both subsequently receive a second
 attractive force from the white particles 
 that sends them into new (larger) epicycles. Finally, when
 our two black particles return to their initial radius (panel 4),
 they are
 azimuthally closer to each other than when they began (in panel
 1). Their gravitational attraction is now stronger and sends them
 into new epicycles of greater amplitude, and the process runs away.
Thus the instability relies on a sequence of
gravitational encounters between neighbouring elements; and these act
as an `epicyclic amplifier'. 

The amplifier we describe here should be contrasted with the `propeller
and frog' resonance (Pan and Chiang 2010). In the set-up of Pan and
Chiang, two `particles' are held fixed ahead of and behind
 a third particle. This third particle, when displaced,
feels the gravitational attraction of the other two, but does not act
back upon them. Consequently, it undergoes a
periodic epicyclic-like motion of a steady amplitude. However, as soon
as we let the third particle react back on its two chaperones, an
instability of the type we describe above should occur. Recently
this seems to have been observed (Pan and Chiang 2012).

\subsection{A gravitational analogue 
of the MRI}

Before we move on to the nonlinear development of the gravitational
instability,
it is of interest to briefly examine a different but related model of
instability, in which the dynamics of the incompressible MRI are reproduced in
most of their details.

Consider, instead of a stream of particles extended along the
azimuthal direction, a \emph{vertical} line of particles located at a single
radius and azimuth: $x_n=y_n=0$. The particles are equally
spaced in $z$. However, in order to keep this configuration in equilibrium it
is necessary to `switch-off' the vertical tidal force of the central
object in Eq.~\eqref{nleq3}. It is also necessary to assume that the vertical line of
particles is extremely long and that we study only particles at the
midplane far from the unbalanced ends. This `equilibrium' may then be
written as
$$ x_n=0,\quad y_n=0, \quad z_n= hn, \qquad \forall n\in \mathbb{Z}. $$
As before, we consider only linear planar disturbances to this basic state,
and we describe this perturbation by 
$$(x_n',\,y_n',\,0) = (X,\,Y,\,0)\,\text{e}^{st + nK\text{i}}.$$
Its linearised equations read
\begin{align} \label{MRI1}
&s^2 X - 2\Omega s Y - 3 \Omega^2 X = -\frac{Gm}{h^3}F\,X, \\
& s^2 Y + 2\Omega s X = -\frac{Gm}{h^3} F\,Y, \label{MRI2}
\end{align}
where $F$ is given earlier in Eq.~\eqref{Force}.

Equations \eqref{MRI1}-\eqref{MRI2}
 describe two bodies connected by a
spring, with a single spring constant $(Gm/h^3)F(K)$. The linear dynamics of the
incompressible MRI can be described in precisely the same way if $v_A^2
k^2$ is substituted for the spring constant (Balbus 2003).
 Under this substitution,
the dispersion relations of the MRI and gravitational instability
  are almost identical!
 The only difference comes from $F$'s
dependence on wavenumber, which is more complicated than $k^2$. Yet
like $k^2$, it is an increasing function and goes to
0 as $k\to 0$. 

So 
in this toy problem we have managed to reproduce the MRI dynamics by
letting the gravitational force between particles
 do the job of magnetic tension in the MRI. And by extending
 the particles in the $z$-direction but suppressing their vertical
 motion we can nullify gravity's tendency to clump the particles (the
 complicating process in Section 2.4). Particles at different $z$
are drawn apart from each
 other in the plane; but these displacements engender
 planar gravitational torques that
 transport angular momentum from particles at smaller radii to
 particles at larger radii. Particles then drift further apart,
 angular momentum is further exchanged, and the process runs away.
 The system develops a vertical sequence of planar jets,
 or `channel flows', as in the MRI (Goodman and Xu 1984).
 Unlike the 
MRI, however, these flows are not nonlinear solutions to the equations
of motion.

\section{Nonlinear and collisional evolution}

In this section we present three-dimensional N-body simulations of the nonlinear and
collisional evolution of the gravitational instabilities discussed. 
The equations \eqref{nleq1}-\eqref{nleq3} are numerically
 evolved forward in
time in a corotating local frame, which imposes a finite
periodicity in the azimuthal ($y$) direction, but has no radial
boundaries. Unlike the shearing box, particles cannot leave the box
through one radial boundary and reappear through the other.
The simulations can show
how the saturation of the instability depends on the governing
parameters of the system, and if the eventual outcome of the clumping
form of the instability differs from the outcome of its oscillatory form.

\subsection{Parameters}

In addition to the dimensionless parameter $g$, which governs the
linear stage, N-body simulations introduce three more parameters
associated with the collisional energy losses, the particle diameter, and
the number of particles $N$ in the box. We, however, have verified
that our results do not depend on $N$, which has no direct physical
meaning. 
Particles are assumed to be
non-spinning inelastic spheres endowed with a constant normal coefficient of
restitution, $\epsilon$. Therefore
\begin{equation}
 v_n' = -\epsilon\, v_n,
\end{equation}
where $v_n$ and $v_n'$ denote
 the normal relative velocities of two colliding
particles before and after the collision respectively. 

The particles' diameter we
denote by $d$, which suggests a dimensionless parameter analogous to
$g$. For consistency with previous work (Ohtsuki 1993, Canup
and Esposito 1995), we use the ratio
\begin{equation}
r_p = \frac{d}{r_\text{H}} = \left(\frac{2G\,m}{3d^3\,\Omega^2}\right)^{-1/3},
\end{equation}
where $r_\text{H}$ is the mutual Hill radius of the particles.
 The parameter
$r_p$  may be recast as
\begin{equation}
r_p = 12^{1/3}\left(\frac{\rho_c}{\rho_p}\right)^{1/3}\left(\frac{R_c}{R_0}\right),
\end{equation}
where $\rho_p$ is the mass density of a particle, $\rho_c$ is the mass
density of the central body, with $R_c$ denoting its radius. In the
case of Saturn's rings, it
follows that $r_p<1$ at radii greater than roughly 125,000 km, if we
take $\rho_p$ to be that of crystalline water ice (900 g/m$^3$). At the radius of
the F-ring, $r_p$ falls to 0.89.

 Finally we can derive the ratio of
the particle radius and initial particle spacing from $d/h= 0.873\, r_p\,g^{1/3}$.
Because we must have $d<h$
 (or else particles overlap), there is a restriction on the size of
 $r_p$ given $g$, which is easy to calculate:
$r_p < 1.145\,g^{-1/3}$.

\subsection{Clustering criteria}
 
The key to the long-term evolution of the ring is the capacity of ring
particles to form long-lived aggregates. In this subsection we discuss
a number of simple criteria that might help us understand the
simulation outcomes.

Initially we consider whether simple aggregates of 
two particles in contact can resist tidal disruption. The simplest
case is that of an aggregate in synchronous rotation, which appears as a
non-rotating aggregate 
in the frame of reference used in this paper. The
particle locations are given by $y_1=y_2=0$ and $x_1=-d/2$ and $x_2=d/2$.
By testing the limiting case of no contact force, Eqs \eqref{nleq1}
and \eqref{nleq2}
tell us that
this configuration is an equilibrium solution if
\begin{equation} \label{clust1}
r_p \leq 1,
\end{equation}
(Weidenschilling et al.~1984, Longaretti 1989).

This `force' condition provides a clear 
criterion, but it applies only when two particles are 
touching in this special configuration. How particles ever
reach such an arrangement is not addressed: they
may not be all that likely in real collisional systems.
Consequently, force criteria may
overestimate clump formation. Another approach is taken by
 Ohtsuki (1993) and Canup and Esposito (1995) who examine the
 energetics of a collision via the Jacobi integral and 
compute a condition for accretion during a binary collision. These criteria relate $r_p$ and
 $\epsilon$. Canup and Esposito find that when
\begin{equation} \label{clust4}
\epsilon < \sqrt{\frac{v_e^2 + (2/3) r_p^2-9}{v_e^2+c^2 + (2/3)r_p^2}},
\end{equation}
colliding particles should aggregate. Here
 $v_e$ 
is escape velocity and $c$ is the velocity dispersion of
the system. Both are scaled by $r_\text{H}\Omega$, and so $v_e=
\sqrt{6/r_p}$.
 If we assume that $c\sim v_e$ in equilibrium, then we have
a relationship between $\epsilon$ and $r_p$ that tells us whether a
typical collision results in aggregation, and hence whether our
system has a tendency to become clumpy. In the limit
of perfectly dissipative collisions ($\epsilon=0$), which is
the most conducive
to aggregate formation, the criterion becomes
\begin{equation} \label{clust5}
r_p \leq 0.691,
\end{equation}
significantly less favourable than \eqref{clust1}.

These criteria hold only for binary aggregates or binary collisions;
but if a binary (or larger) 
clump has already formed then additional particles can attach themselves 
to it with
greater ease (Weidenschilling et al.~1984). It follows that
 if criterion \eqref{clust1} is satisfied and yet
\eqref{clust4} is not satisfied,
 a rare event may
occur whereby a long-lived aggregate forms and then successfully
absorbs other particles, thereby growing larger and larger. 
For fixed
$\epsilon$, it follows
that an intermediate regime may
exist, with $r_p$ lying between the limits obtained from \eqref{clust4} 
and \eqref{clust1} (Salo 1995).
 For $\epsilon=0$, this would be $0.691<r_p<1$.
In this regime long-lived clumps
do not form in general --- almost all collisions do not result in capture
---
 but given sufficient time a rare particle encounter occurs that
does result in an aggregate. This aggregate may then grow
successfully, an embedded blob in an otherwise monodisperse ring.

\subsection{Numerical method}

Using the collisional N-body code REBOUND (Rein and Liu 2012) we 
study the gravitational instability by directly evolving
 Eqs.~\eqref{nleq1}-\eqref{nleq3}. 
Because of the small number of particles, we can use direct 
summation to accurately calculate self-gravity and do not have
 to use a tree or FFT-based gravity solver.

We use the mixed-variable symplectic epicycle integrator SEI
 (Rein and Tremaine 2011) which is ideally suited for this study.
A symplectic integrator does not introduce artificial trends
 in formally conserved quantities.
The mixed-variable integrator gives a large accuracy gain
 when the particle motion is dominated by epicyclic motion.
Because of these properties, we have numerically converged 
results in most simulations using a time-step as large as 
one tenth of the dynamical time, $dt=10^{-1}\Omega^{-1}$.

Usually, one employs several ghost boxes in local N-body
 simulations to ensure that there is no special place in the shearing sheet.
The more ghost boxes are used, the better. 
Here, we find it is sufficient to use
 only one ghost box in both the positive as
 well as the negative $y$ direction. 
There are no ghost boxes in the $x$ direction, since we 
are dealing with a string of particles. Thus each particle feels the
attraction of $3N-1$ other particles.

We pre-calculate the residual force error $E_i$ in the calculation of
$f_y$ comparing the initial numerical setup to a truly infinite string
 of particles with perfect periodic spacing $h$, cf.\
 Eq.~\eqref{Eqm}. This is similar to the idea of Ewald summation, and
the error can be written down explicitly in terms of the 
first derivative of the digamma function
\begin{equation}
E_i = \frac{Gm}{h^2} \left(\psi'(2N-i)-\psi'(1+i+N)\right),
\end{equation}
where $i$ is the index of the particle running from $0$ 
(left) to $N-1$ (right), and a prime denotes differentiation.
This error is then subtracted from the numerically 
calculated forces at every time-step.
Using this trick, artificial effects are minimised
 at the box boundaries even when the instability is
 followed over many dynamical timescales and grows 
by many orders of magnitude.

We use an instantaneous collision model. Hence multiple collisions during one
time-step may not be treated correctly. Furthermore, we use a minimal impact
velocity ($0.05\,d\Omega$ ) so as to avoid overlapping particles when aggregates
form. 
This velocity scale
is set much smaller than the velocity dispersion and so should not affect the outcome.

\subsection{Linear regime}

\begin{figure}
\scalebox{1.0}{\includegraphics{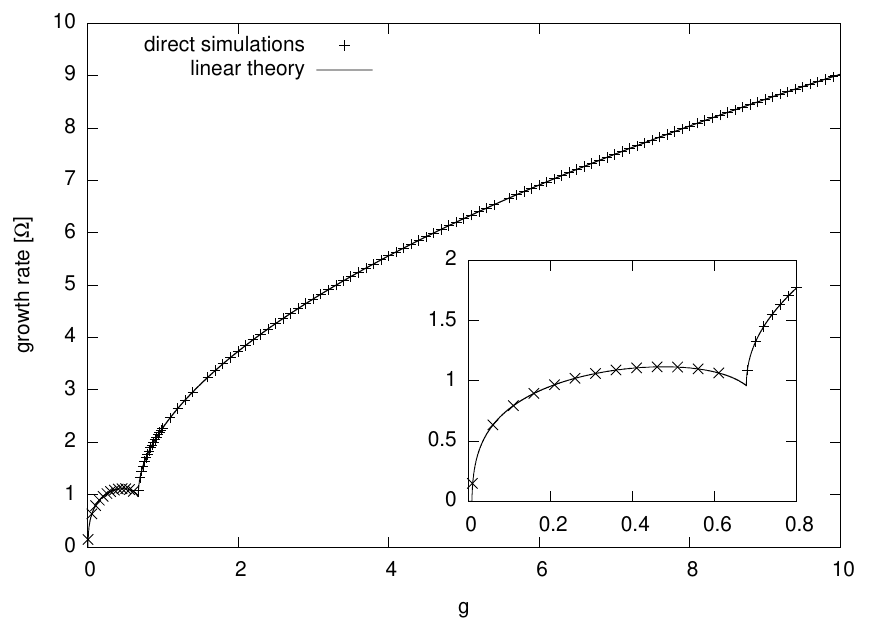}}
\caption{Largest growth rate of the system in numerical simulations
 and linear theory as a function of the parameter $g$. Note that the
 detailed structure at small $g$ is captured perfectly. 
 Here $N=200$. \label{fig:rebound:growthrate}}
\end{figure}

As a numerical check, the analytic growth rate 
given by Eq.~\eqref{disp} is verified using the N-body code. 
We take $N=200$ particles initially 
lined up in a string and simulate their early evolution.
The particles' location and velocity are
 perturbed with the unstable eigenfunction of fastest growth. 
The initial perturbations possess amplitudes of $10^{-7}$ and we stop
the evolution when they have
grown by 2 orders of magnitude. 

In Fig.\ref{fig:rebound:growthrate} we plot
 the largest growth rate as a function of the dimensionless parameter $g$. 
We also show the analytic result from linear theory. 
One can see that the agreement is excellent, which provides a useful 
validation of the numerical integrator, and also a check on the
analytic theory.

\subsection{Nonlinear evolution}

\begin{figure*}
\centering 
\scalebox{1.05}{\includegraphics{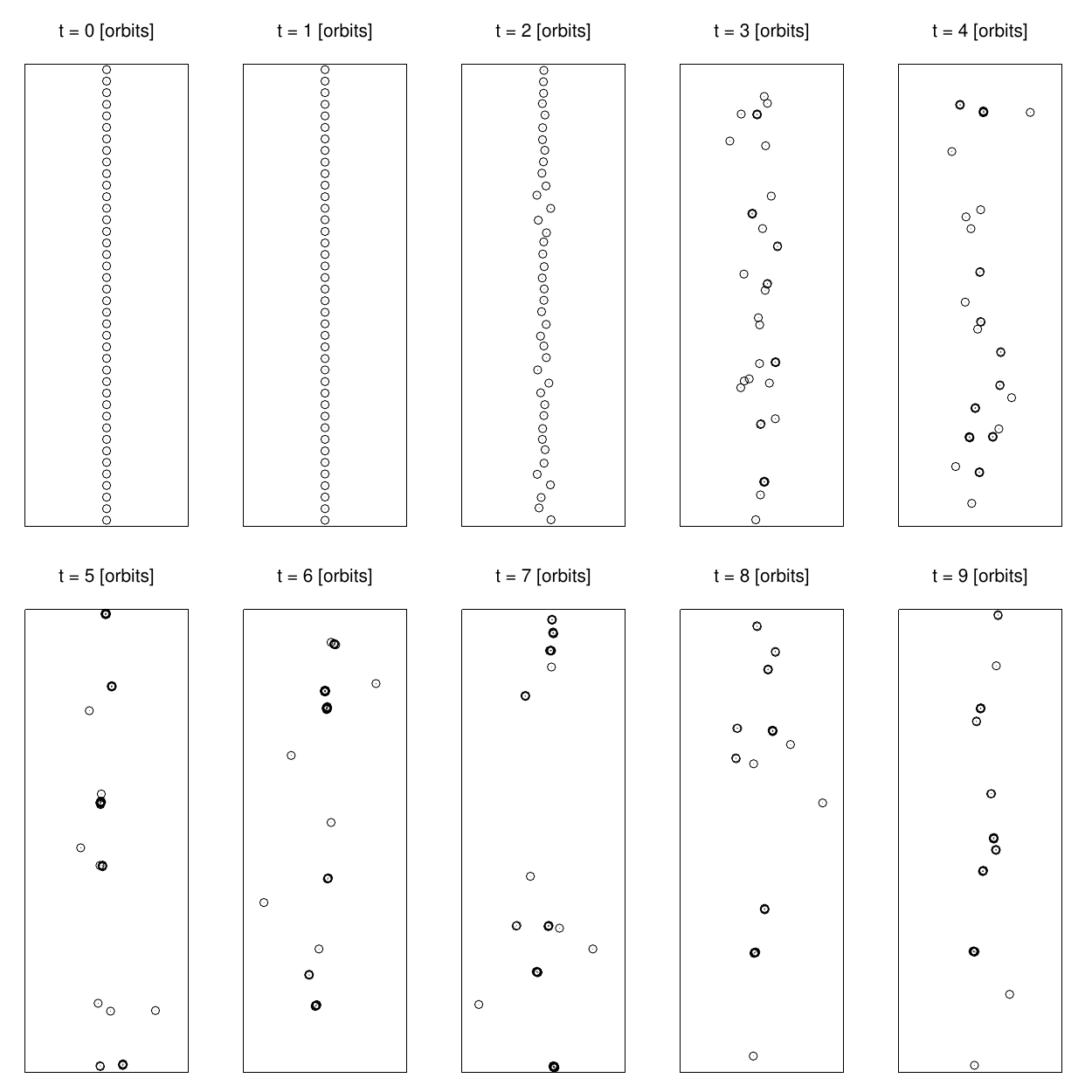}}
\caption{Snapshots of a simulation of gravitational instability.  
Parameters are: $g=0.1$, $r_p=0.2$, $\epsilon=0.5$,
$N=200$. Note that only a central portion of the computational domain
is shown. Particle sizes have been
inflated to aid their visualisation; hence particle aggregates appear
as overlapping particles.
The simulation is initialised by small amplitude random noise. \label{clumping}}
\end{figure*}

\begin{figure*}
\centering 
\scalebox{1.05}{\includegraphics{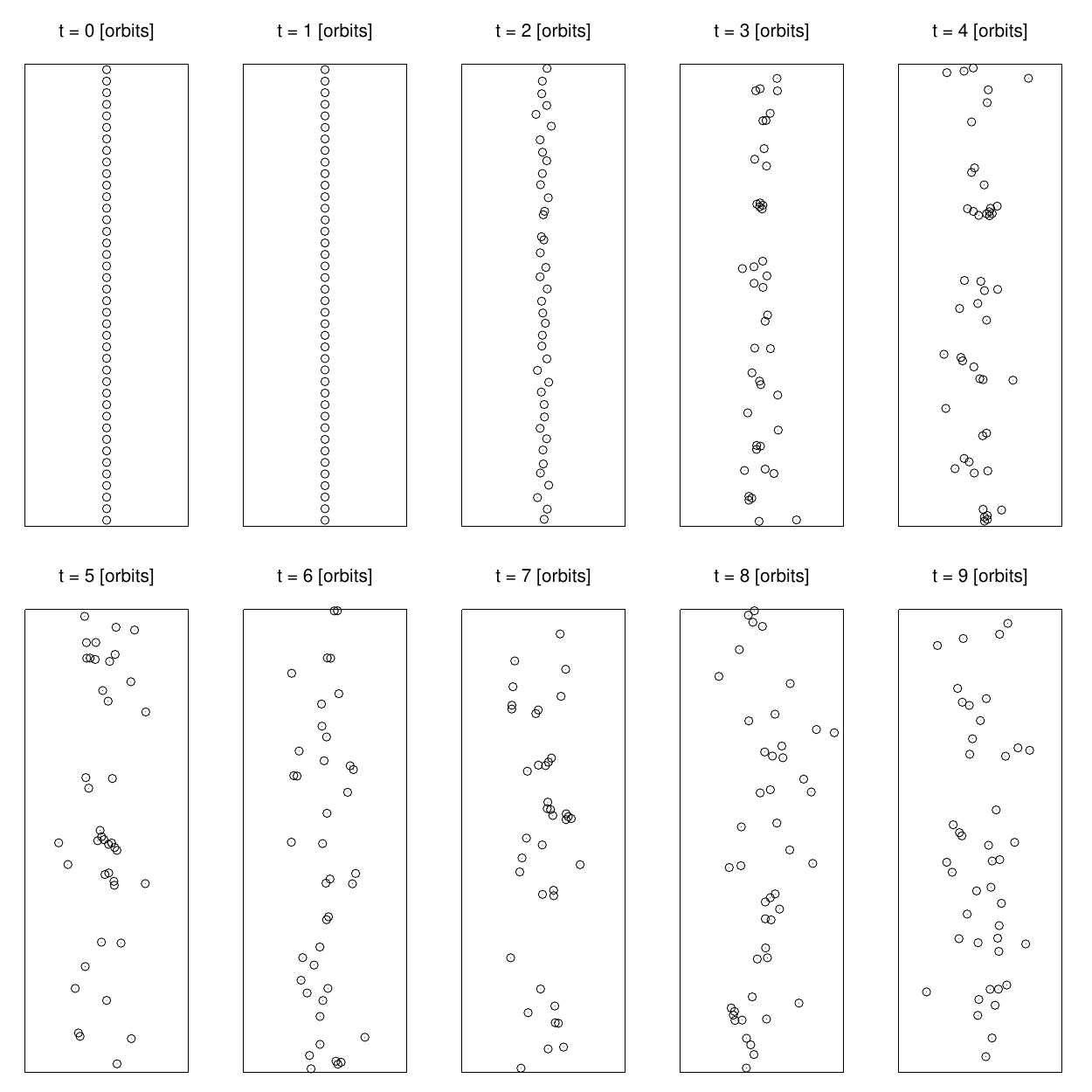}}
\caption{As in Fig.~\ref{clumping}, but with parameters 
$g=0.1$, $r_p=0.8$, $\epsilon=0.5$,
$N=200$.  
\label{oscill}.}
\end{figure*}

Long time simulations were ran for a variety of parameters, though
first we
concentrate on the following four parameter sets presented in Table I.
\begin{table}
\begin{center}
  \begin{tabular}{c | c | c | c | c | c }
    \hline
    Case & $g$ & $r_p$ & Instability & Aggregation & $d/h$ \\ 
    \hline
    (i) & 0.1 & 0.2 & Oscill. & Likely & 0.081 \\ \hline
    (ii) & 0.1 & 0.8 & Oscill. & Unlikely & 0.32 \\ \hline
    (iii)  & 1 & 0.2 & Monot. & Likely & 0.17 \\ \hline
    (iv) & 1 & 0.8 & Monot. & Unlikely & 0.7  \\ \hline     
    \hline
  \end{tabular}
\caption{Parameter sets for the four runs discussed in detail in Section 3.5}
\end{center}
\end{table}
In the table, `Oscill./Monot.~' refers to the form of the initial
instability, oscillatory or monotonic (determined by $g$), and `likely'
or `unlikely' aggregation refers to
the likelihood of clumping as decided by \eqref{clust4}.
In all cases we set $\epsilon=0.5$ and $N=200$. Note
 that the $r_p=0.8$ cases fall within the
`intermediate regime' discussed earlier: aggregates can still exist 
in principle.
These parameter choices hence let us observe the potentially different outcomes
of the two forms of gravitational instability, while also testing the
clustering criteria of Section 3.2.  

We find that on medium to long times the simulations yield two
typical evolution tracks depending on the parameter $r_p$. It
appears that after a few orbits the ensuing disordered state
 retains little memory of the linear instability that gave rise
 to it. We take simulations (i) and (ii) as examples of these two
 tracks.

First we treat case (i), as it is the simplest. These parameters yield
oscillatory instability and physical collisions that usually lead to
capture.
 In Fig.~\ref{clumping}
screenshots of the evolution are presented corresponding to
different times. In order to better visualise the results, 
these images only represent the central portion of
the computational domain, comprising initially of 40 particles. 
Particles are represented by black circles and
 their radius has been inflated so they are more easily seen. 
In panel 1, the equilibrium set-up is shown, while in
panel 2, the system is on the verge of departing from the linear regime
of the oscillatory
instability. However, the perturbations are still too small to be
easily observed. Panel 3 shows the outcome of the first round of collisions.
These initial collisions induce accretion and
precipitate disordered non-planar motion in the resulting aggregates. Gravitational
encounters liberate orbital shear energy and the ring `warms'
up with a velocity dispersion $\sim v_e$; in addition, angular
momentum is transported and the ring spreads, its width eventually $\propto t^p$,
for $p$ close to $1/2$. Particles and
particle aggregates continue accreting through physical collisions, in
accordance with the criterion \eqref{clust4}, and after 8 orbits
the number of bodies has decreased significantly. Because we have
inflated the particle radiuses in the images, the 
aggregates appear as overlapping circles.

The evolution of case (ii) is represented by eight screenshots in
Fig.~\ref{oscill}. Unlike case (i), the system does not steadily
collapse into a smaller set of aggregates. Instead it swiftly degenerates
into a disordered spreading state characterised by a velocity
dispersion, as before, of order $v_e$ (consistent with Ohtsuki 1999
and Ohtsuki and
Emori 2000). The eight panels clearly
describe this spreading. However, what they also show is the
continuous generation and dissolution of small particle
aggregates. Particles come together and then are disrupted by
tidal forces, spin, or forceful collisions with other particles. These
temporary aggregates resemble the dynamic ephemeral bodies (DEBs)
proposed by Weidenschilling et al.\ (1984).
 Alternatively, we may
think of them as analogues of the gravity wakes exhibited by optically
thicker rings (Salo 1992).  We have found that some
aggregates persist for the length of the simulation.
These bodies may have become
sufficiently large to pass the
tidal disruption threshold for these parameters. Simulations were also
run for $r_p=0.85,\,0.9,$ and 1 and long-lived clusters failed to
appear, though occasional two-particle aggregates were spotted. 

 These runs confirm an intermediate regime of $r_p$
 between continuous aggregation and no aggregation at all. This regime
 is characterised, in particular, by temporary aggregates and
 by the potential formation of persistent growing clusters that resist
 disruption amidst the surrounding melee.

\subsection{Clumping criterion}

In order to measure the efficiency of clustering in detail we
conducted a parameter survey of $\epsilon$ and $r_p$. 
We employ a convenient `one-dimensional' measure of the system's degree of
clustering which we now describe. 
Particles are sorted along the azimuthal direction
and the azimuthal distances between neighbouring particles $\Delta y$
calculated. (Their relative radial distances are discarded.)  
The mean of the resulting distance distribution $\langle \Delta y \rangle$
stays constant throughout a simulation and
is equal to $h$. However, its standard deviation will evolve over time. 
We average this standard deviation over the length of the run and
scale it by $h$; the
final averaged quantity we denote by $q$ and associate it with the
degree of aggregation throughout the run.
Note that $q=0$ in the perfectly ordered equilibrium state that we
start with, while $q=1$ if the particles are distributed entirely
randomly. In the latter case, $\Delta y$ exhibits an exponential distribution with
scale parameter $h$. 
Aggregation corresponds to $q>1$, with $q$ taking its
maximum value $q\approx \sqrt{N}$ when all the particles lie in a
single cluster. In our simulations, we find that $q$ never exceeds
about 6. The one-dimensional measure $q$ is well-suited to a narrow
spreading ring, and is also simpler to implement than the two-dimensional Renyi entropy employed in
Karjalainen and Salo (2004). Like the
Renyi entropy $q$ cannot distinguish between the continuous formation
of many temporary clusters and permanent aggregates.

\begin{figure}
\scalebox{0.85}{\includegraphics{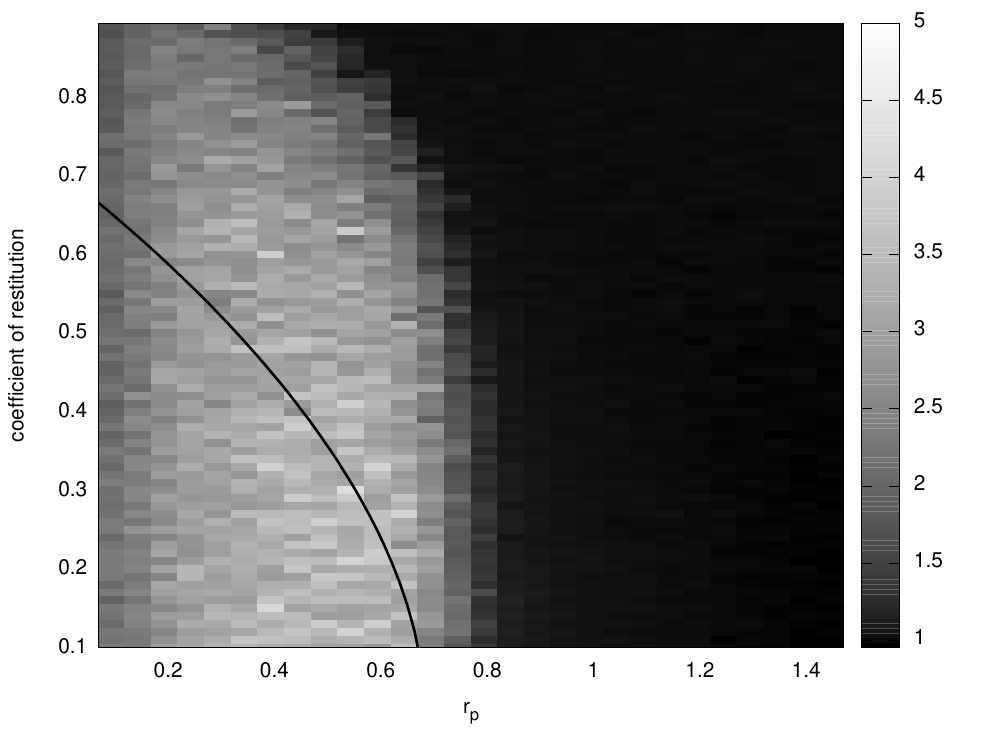}}
\caption{The measure of clumpiness $q$ as a function of $\epsilon$ and
  $r_p$. Superimposed on the greyscale plot
is the Canup and Esposito criterion (1995) for gravitational 
capture after a binary collision (the black line).
 The other parameters are $g=0.1$ and
$N=200$. \label{dispersion}}
\end{figure}

Figure \ref{dispersion} plots $q$ in greyscale as a function
of $\epsilon$ and $r_p$. It summarises 2240 separate simulations each 
run up to 200 orbits. The figure
illustrates clearly the hard boundary at $r_p=1$;
for larger $r_p$, we find $q=1$ and the particles are distributed
randomly with little aggregation. Moreover, our results show that
aggregation is insignificant until at least $r_p\approx 0.8$, though this limit
decreases for larger $\epsilon$. The $q$ value jumps
abruptly near this critical $r_p$, from slightly greater than 1
(little accumulation),
 to
$q\approx 5$, which we associate with the onset of continuous
accretion. In addition we have overplotted the Canup and
Esposito criterion (1995) given in Eq.~\eqref{clust4}, which appears
to underestimate the prevalence of clustering, especially for higher $\epsilon$.

 At small $r_p$ it appears
that $q$ decreases slightly. This is perhaps due to the fact that in this
regime the particles are so small that they rarely collide with
each other. Extremely long time integrations are hence required to
reach the systems' accretion time-scales. Our simulations, which were run
up to 200 orbits, were perhaps too short. It is also possible that
accretion is suppressed in hot
rings composed of very small particles.

The numerical results are compatible with self-gravitating simulations
of broad rings
(Salo 1995, Karjalainen and Salo 2004), which also
generate long-lived aggregates. However, 
the optical
depth is much greater in those simulations, and they are
significantly perturbed by gravity wakes --- conditions that make
aggregate formation more likely and more rapid than in our work.
 In particular, 
Karjalainen and Salo (2004) find that when $\tau=0.25$ continuous
accretion occurs for $r_p \lesssim 0.87$ if $\epsilon=0.1$, and for
$r_p\lesssim 0.85$ if $\epsilon = 0.5$. These critical values are
larger than ours, but at
lower $\tau$ their critical $r_p$ approaches 0.8, which is closer to
our critical value. It should be noted that these low $\tau$ runs
 were undertaken with a
varying coefficient of restitution.

\section{Conclusion}

In this paper we have studied the gravitational and collisional
dynamics of a stream of co-orbital particles, plotting the course of
the system's evolution from initial
instability to its nonlinear saturation. Gravitational instability
attacks the equilibrium in one of two ways: as either a
monotonic clumping or as growing epicyclic
motions. The latter's mechanism is particularly interesting, as it consists
of the interaction of gravitational attraction and angular momentum
exchange. In this way it shares some properties with other disc
instabilities, such as the Papaloizou-Pringle instability and the
MRI.

The nonlinear collisional evolution of the instabilities are tracked
with N-body simulations. These show that the final saturated state is
relatively insensitive to the form of the initial instability and is
instead controlled by the inelasticity of collisions (measured by
$\epsilon$, the coefficient of restitution) and the ratio of particle
diameter to the mutual Hill radius ($r_p$). The system tends to move towards
 three regimes: (a) for $r_p \gtrsim 0.83$ a hot disordered flow ensues
 with little or no clumping; (b) for $r_p$ less than 0.83 but greater
 than a second critical value $r_c(\epsilon)$, the system
 continuously forms short-lived small aggregates and occasionally a
 permanent cluster; (c) when $r_p<r_c$ most collision result in
 accretion and the system accumulates into permanent and
 continuously growing conglomerates.

Regime (b) is especially relevant to conditions in the F-ring
of Saturn, where the $r_p$ parameter is close to 0.83. Our
simulations, like Salo (1995) and Karjalainen and Salo (2004), are
hence consistent with observations of embedded large bodies in the
F-ring,
and theories that attribute their prevalence
to a size distribution dynamics comprising gravitational
aggregation, and tidal and collisional disruption (Barbara and
Esposito 2002, Esposito et al.~2011). Our simulations directly
 animate these processes
from first principles, though the real system exhibits many more
physical effects than our model. These include `stirring' by moons,
adhesive surface forces, and a size distribution.

Finally, the linear instabilities presented here may have analogues in
dense narrow rings that exhibit orbital shear (Goodman and Narayan
1988, Papaloizou and Lin 1989). Gravitationally
 unstable streams of
material torn from tidally disrupted satellites may be the progenitors
of these narrow rings, including the Uranian $\epsilon$-ring (Leinhardt
et al.~2012). Allowing for the modifications wrought by shear, the
basic mechanism of instability should be much the same, combining
gravitational clumping and angular momentum exchange. The simple
system we study here, which permits
 us to isolate and understand the salient physics,
 can then provide an inroad into 
the more complicated dynamics of the confined shearing system.

\section*{Acknowledgements}

The authors would like to thank the reviewer, Heikki Salo, for
a thoughtful review which helped improve the quality of the paper.
Henrik Latter and Gordon Ogilvie acknowledge funding
 from STFC grant ST/G002584/1. 
Hanno Rein was supported by the Institute 
for Advanced Study and the NSF grant AST-0807444.
Simulations in this paper made use of
 the collisional N-body code REBOUND which 
can be downloaded freely at \url{http://github.com/hannorein/rebound}.

\end{document}